\documentclass{nature}
\usepackage{amsfonts, amsmath, amssymb, amsthm, array, eucal, mathtools, xfrac}
\usepackage{booktabs, array, color}
\newlength\fwidth
\newlength\fheight
\usepackage{graphicx}
\usepackage{siunitx}
\usepackage{enumitem}
\usepackage{epstopdf}

\title{Shear-induced gelation of self-yielding active networks}

\author{David A. Gagnon,$^{1,2,*}$ Claudia Dessi,$^{1,2,*}$ John P. Berezney,$^3$ Daniel T.-N. Chen,$^3$ Remi Boros,$^4$\\ Zvonimir Dogic,$^{3,4}$ \& Daniel L. Blair$^{1,2}$}

\begin{document}

\maketitle

\begin{affiliations}
 \item Department of Physics, Georgetown University, 3700 O Street NW, Washington, DC 20057, USA
 \item Institute for Soft Matter Synthesis \& Metrology, Georgetown University, 3700 O Street NW, Washington, DC 20057, USA
 \item Department of Physics, Brandeis University, Waltham, MA 02453, USA 
 \item Department of Physics, University of California Santa Barbara, Santa Barbara, CA 93106, USA
 $^*$These authors contributed equally to this work
\end{affiliations}

\begin{abstract}
Molecular-motor generated active stresses drive the cytoskeleton away from equilibrium, endowing it with tunable mechanical properties that are essential for diverse functions such as cell division and motility\cite{RN73,RN6,RN72,RN3,RN5}. Designing analogous biomimetic systems is a key prerequisite for creating active matter that can emulate cellular functions\cite{RN1,RN71}. These long-term goals requires understanding of how motor-generated stresses tune the mechanics of filamentous networks\cite{RN67,RN4,RN2,RN74,RN19}. In microtubule-based active matter, kinesin motors generate extensile motion that leads to persistent breaking and reforming of the network links\cite{Henkin2014}. We study how such microscopic dynamics modifies the network's mechanical properties, uncovering that the network viscosity first increases with the imposed shear rate before transitioning back to a low-viscosity state. The non-monotonic shear-dependent viscosity can be controlled by tuning the speed of molecular motors. A two-state phenomenological model that incorporates liquid- and solid-like elements quantitatively relates the non-monotonic shear-rate-dependent viscosity to locally-measured flows. These studies show that rheology of extensile networks are different from previously studied active gels\cite{RN69}, where contractility enhances mechanical stiffness. Moreover, the flow induced gelation is not captured by continuum models of hydrodynamically interacting swimmers\cite{PhysRevE.83.041910, PhysRevLett.92.118101,Gachelin2013,RevModPhys.85.1143,Saintillan2018,Sokolov2009,Rafai2010,Saintillan2010}. Observation of activity-dependent viscoelasticity necessitates the development of models for self-yielding of soft active solids whose intrinsic active stresses fluidize or stiffen the network.
\end{abstract}

\begin{figure}[ht]
\centerline{\includegraphics[width=0.8\textwidth]{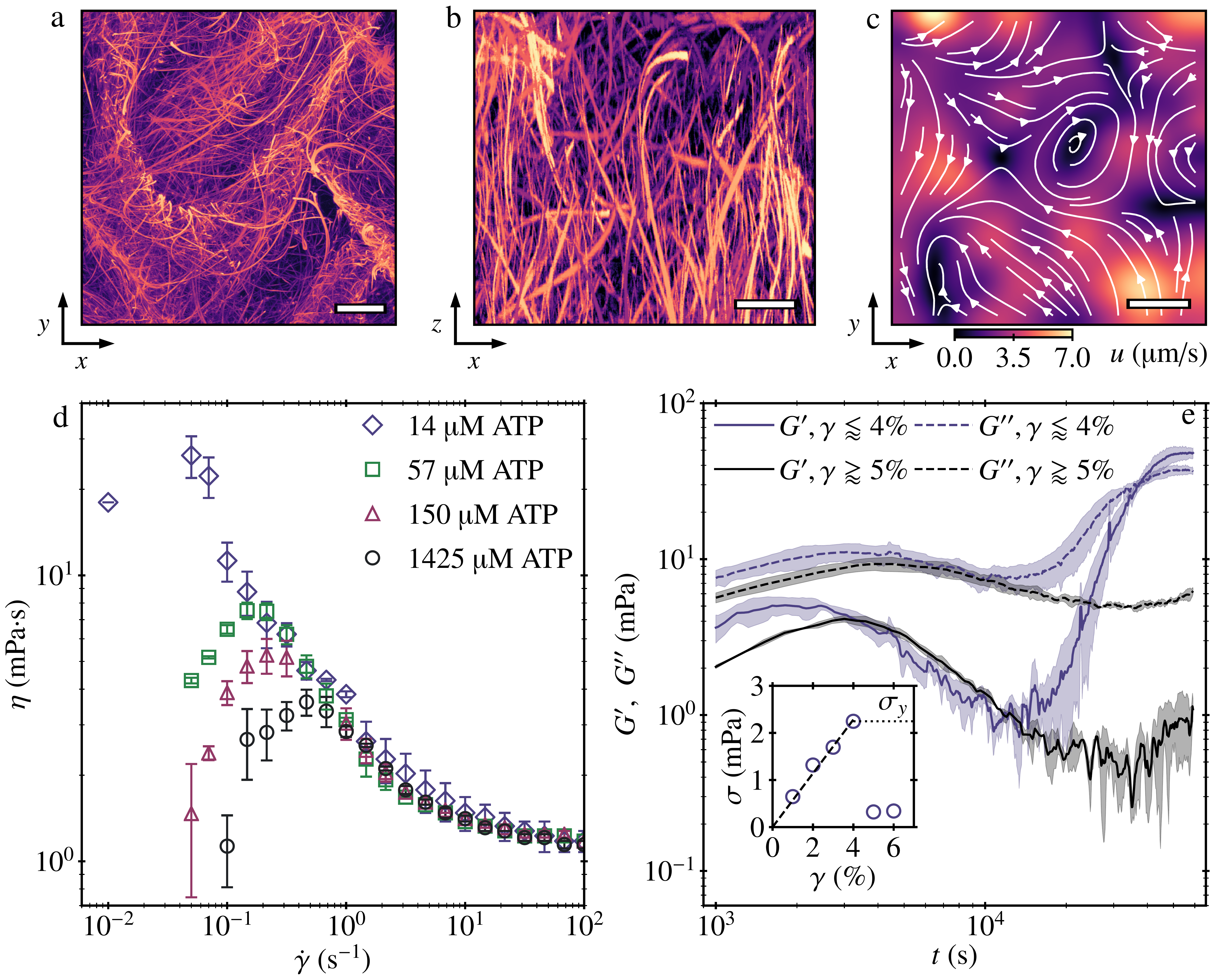}}
\caption{{\bf Structure, dynamics and rheology active gels:} {\bf a-b} Confocal fluorescence images of active network in the {\em x,y}- (velocity, vorticity) plane and the {\em x,z} (velocity, gradient) plane as the ATP concentration is depleted. Images represent $100$~\si{\micro}m maximum projections. {\bf c} Velocity field of autonomous flows and streamlines in the flow-vorticity plane with no applied shear. {\bf d} Apparent viscosity $\eta$ as a function of shear rate $\dot{\gamma}$; error bars reflect standard error across multiple samples ($N=3$). {\bf e} Storage $G^{\prime}$ and loss $G^{\prime \prime}$ moduli of network initially with $c_\mathrm{ATP} = 14$ \si{\micro}M suspension as ATP. Frequency is fixed $\omega = 1.0$rad s$^{-1}$. The light colored curves indicate the oscillatory tests below yield stress $\sigma_y$, and the dark curves are above $\sigma_y$. Shading indicates the standard deviation for three experiments. \textit{Inset:} Stress measurements from a strain sweep provide an approximate yield stress $\sigma_y=2.2$~mPa. Scale bars are 50~\si{\micro}m. \label{fig1} }
\end{figure}

We study the mechanical properties of isotropic active networks composed of filamentous microtubules (MTs), kinesin motor complexes, and a depleting  polymer that bundles MTs\cite{RN39}. Fueled by adenosine triphosphate (ATP), kinesin clusters bind to and repeatedly step along multiple MTs, producing predominantly extensile bundle motion, which in turn generates bundle buckling, fraying, and disintegration (Supplemental Material, Movies S1-S3). Disintegrated bundles rapidly reincorporate into the network, forming new filamentous paths that continuously reconfigure the structure of the percolating network (Fig.~\ref{fig1}{\bf{a,b}}). Several features make  microtubule-based materials an ideal model system of active matter. First, the rate of network rearrangements can be tuned by controlling ATP concentration, which determines the stepping speed of kinesin motors\cite{RN75}. Second, the efficient kinesin motors can sustain steady-state active dynamics and associated autonomous flows for hours (Fig.~\ref{fig1}{\bf{c}}). Third, MT networks can be assembled on macroscopic, milliliter-scale quantities. However, rheological characterization of these materials is challenging due to a very low MT volume fractions, $\phi = 0.01$ wt\%. To overcome this obstacle we pushed the limits of the instrument sensitivity by designing custom rheological tools (Supp Fig. 1). This advance, when combined with the ability to prepare large (~1 mL) samples that sustain non-equilibrium activity for multiple hours, enabled quantitative rheological characterization. 

Measurements revealed that the sample viscosity $\eta$, exhibits a pronounced non-monotonic dependence on the applied shear rate $\dot\gamma$. For small $\dot\gamma$, the active MTs are shear-thickening, with $\eta$ increasing sharply and reaching a peak, followed by a shear-thinning response (Fig.~\ref{fig1}{\bf d}, Supp Fig. 2). The characteristic shear rate $\dot{\gamma}_c$ corresponds to the peak in the viscosity that separates the shear-thickening from the shear-thinning regime. Repeating these experiments for different ATP concentrations revealed that the internal motor-generated dynamics tunes the network viscosity. In the low-rate shear-thickening regime, changing ATP concentration modified viscosity by up to an order of magnitude. Conversely, in the shear-thinning regime above the critical shear rate $\dot\gamma_c$, the sample viscosity was ATP-independent, with all measurements collapsing on each other. The characteristic shear rate ($\dot{\gamma}_c$)  increased with increasing ATP concentration, while the magnitude of the maximum viscosity simultaneously decreased. 

To explain the non-monotonic viscosity, we note that kinesin clusters have a dual role in the network dynamics by both generating active stresses and by passively cross-linking MTs. Processive kinesin motors take successive $8$~nm steps along a MT backbone, with duration of each step taking tens of microseconds or less\cite{RN23,RN77}. In between steps, motors experience extended milliseconds-long dwell times during which they are statically bound to MTs, and thus acting as passive cross-linkers. Both the concentration of ATP and the magnitude of the applied load determines the duration of the extended dwell times between rapid steps\cite{RN23,RN28}, controlling whether a kinesin motor cluster is an active stress generator or a passive cross-linker. In the absence of ATP, motors permanently link the microtubule network (Supp. Fig. 3{\bf a}). 

To demonstrate the dual role of kinesin clusters, we assembled an active MT network at an intermediate ATP concentration (Movies S2,S3). As ATP was depleted, the dwell times become longer and the dominant role of motor clusters switched from active stress generation to passive cross-linking. Such networks were similar to previously studied  biopolymer networks of cross-linked biofilaments\cite{RN7,RN37}. To quantify how the mechanical properties of the network change during the consumption of ATP, we use oscillatory rheology to continuously measure the elastic $G^\prime$ and loss modulus $G^{\prime\prime}$ at a fixed frequency and amplitude (Fig.~\ref{fig1}{\bf e}). The initial response was viscous dominated ($G^{\prime\prime} > G^{\prime}$), indicating that motor-generated active stresses fluidize the network. As ATP was depleted, the elastic modulus increased, and eventually dominated the loss modulus ($G^{\prime} > G^{\prime\prime}$), confirming the hypothesis that clusters crosslink adjoining microtubule bundles (Supp Fig. 3). Confocal images of the depleted network show that the bundles connect the surfaces of the rheometer, giving rise to an elastic modulus (Fig.\ref{fig1}{\bf b}). We estimate the yield stress of the ATP-depleted network by performing a strain amplitude sweep at a fixed frequency. Above a critical amplitude $\gamma_{\rm{y}}\approx 4\%$, the soft elastic network breaks down at a stress $\sigma_y =2.2$~mPa. The measured stress $\sigma$ increases linearly with the imposed strain, resulting in an elastic modulus of $\sim 60$ Pa, consistent with the elastic modulus of MT network when the ATP is fully depleted (Fig.~\ref{fig1}{\bf e}, Supp Fig. 5).

\begin{figure}[ht]
\centerline{\includegraphics[width=\textwidth]{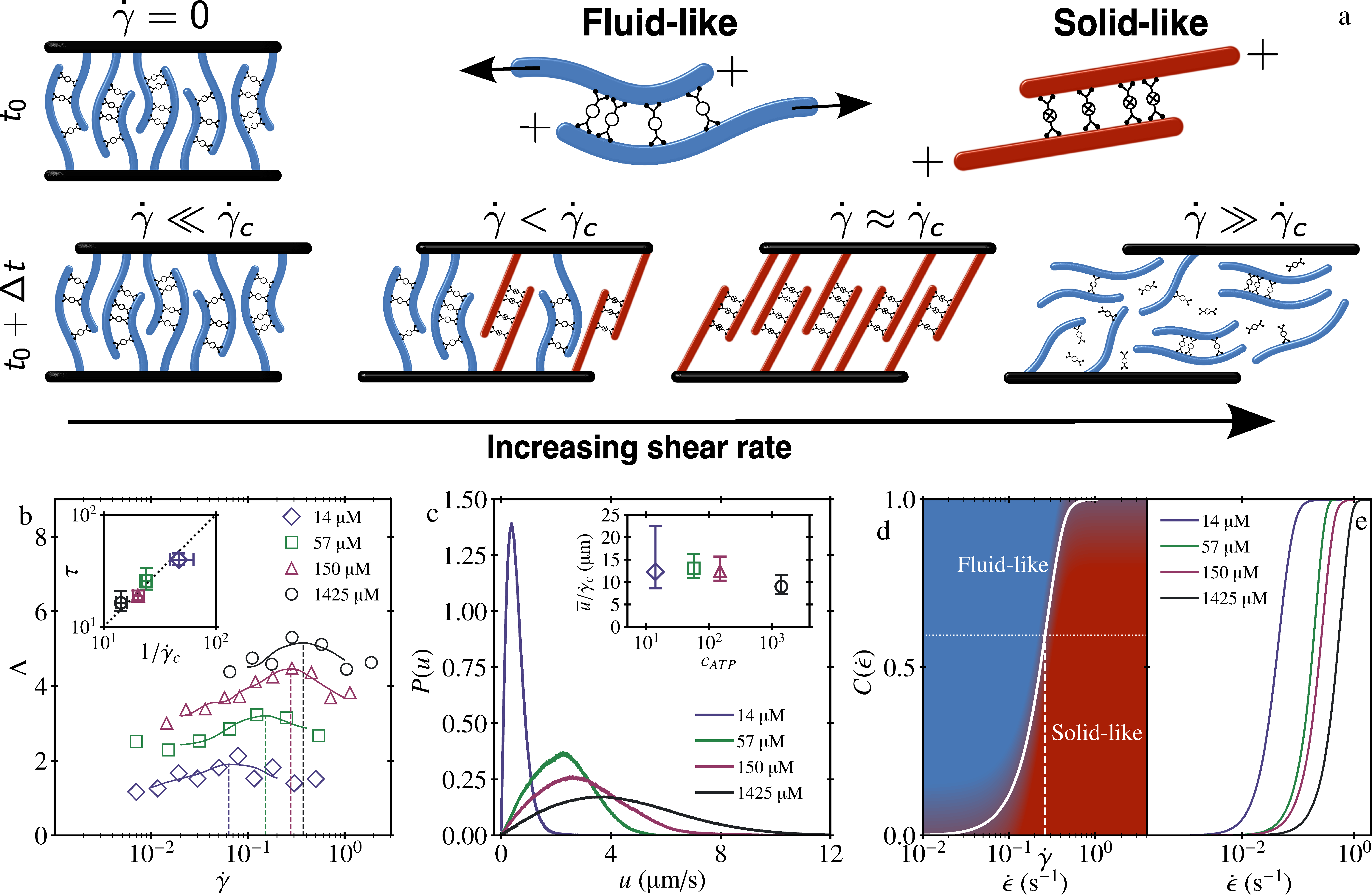}}
\caption{{\bf Microscopic model of active self-yielding solids:} {\bf a} A sheared active network is composed of liquid-like (blue) and solid-like (red) extensile elements. The fraction of each element depends on the speed of intrinsic extension and the magnitude of the applied shear. Close to $\dot\gamma_c$ the speed of the boundary is comparable to the speed of the extending elements, which became taught and contribute to shear stress. Above $\dot\gamma_c$ the extending elements break immediately. {\bf b} Ratio of characteristic length scales in the shear and vorticity directions; a vertical shift is introduced for clarity. \textit{Inset:} The active timescale $\tau$, measured via MT dynamics, is directly correlated to the rheological timescale $1/\dot{\gamma}_c$, dashed line is a fit of slope unity. {\bf c} Probability distribution functions of MT-speeds for different $c_\mathrm{ATP}$.  \textit{Inset:} $\overline{u} / \dot{\gamma}_c$ is independent of $c_{\rm{ATP}}$. {\bf d} The fraction of stiffened elements is set by evaluating the cumulative distribution of sliding rates $C(\dot\epsilon)$ at the shear rate $\dot\gamma$. {\bf e} $C(\dot\epsilon)$ for all $c_{\rm{ATP}}$. \label{fig2}}
\end{figure}

In static situations, thin filaments can carry large tensile loads, but easily buckle under compressive axial stresses. In a dynamical case, where the filament is extending and boundaries are moving, the ability of the filament to sustain loads will depend on the relative speeds of the two dynamical elements.  Motivated by these considerations, we hypothesize that sheared MT networks can be described by two mechanical states, a soft solid-like element with a finite yield stress that coexists with a fluid-like element (Fig.~\ref{fig2}{\bf a}). The relative fraction of these states is set by the two competing rates that characterize the system dynamics; one rate is determined by the externally applied shear, while the other is determined by the ATP-dependent internal rate of extensile sliding. Analogous to the static case mentioned above, network elements that extend at rates faster than the externally-applied deformation self-yield and do not contribute elastically. Conversely, elements whose extensile sliding rates are slower than the externally imposed rates resist the shear and contribute elastically. 

At very low shear rates ($\dot\gamma \ll \dot\gamma_c$) the entire system continuously self-yields and the fluid-like elements determine the mechanical response. As the characteristic shear rate is approached from below ($\dot\gamma \approx \dot\gamma_c$), the imposed rate becomes larger than the intrinsic extensile sliding rate, leading to an increased fraction of elements that become taut and elastically resist the imposed deformation. Once they reach their yield strain, they must break. Simultaneously, the internal activity leads to formation of new extending elements. The continuous breakage and reformation of bundles produces a constant fraction of solid-like bundles that resist shear and thus increase the effective sample viscosity. Such a state is reminiscent of conventional soft-solids; elastic materials that have a high shear viscosity due to the continuous formation and breakage of elastic bonds\cite{annurev.fl.24.010192.000403}. Finally, when $\dot\gamma > \dot\gamma_c$, the imposed shear rate is larger than the extensile sliding rate of all network elements, resulting in their immediate breakage. The model predicts that solid-like elements behave elastically at $\dot\gamma_c$, resulting in an effective increase in stress for $\dot\gamma< \dot\gamma_c$. The network elements break at the yield stress of the inactive MT-gel (Fig.\ref{fig1}{\bf e} inset), determining the rheological response at $\dot\gamma > \dot\gamma_c$. Prior to breaking, taut filaments will microscopically appear solid-like, with increasingly correlated velocities in the shear direction, as has been observed for other yield-stress materials such as dense emulsions\cite{PhysRevLett.120.018001}.

To test our hypothesis, we first quantify how the velocity fields are altered by shear to identify microscopic signatures of shear-induced network solidification. Particle imaging velocimetry reveals the flows fields in the frame of reference where the average flow is subtracted (Supp Fig. 3). From here, we determine the velocity autocorrelation lengths both in the direction parallel to and perpendicular to the flow (Supp Fig. 4). Their ratio, $\Lambda$ is a non-monotonic function of $\dot\gamma$ with an ATP dependent peak (Fig.~\ref{fig2}{\bf b}). The peak in $\Lambda$ defines a timescale $\tau$ that is linearly correlated with independently measured characteristic shear rate $\dot\gamma_c$ where the sample viscosity is maximum  (Fig.~\ref{fig2}{\bf b} inset). As the system becomes more solid-like at $\dot\gamma_c$, the range of the velocity correlations in the direction of shear flow increases as the filaments begin to span the system and align along the principle shear axis\cite{PhysRevLett.120.018001}.  

Next, we compare the external and internal rates to determine the fraction of solid-like elements. Particle imaging velocimetry in the absence of an imposed shear determines the distribution of microscopic strain rates the are generated by the autonomous flows (Fig.~\ref{fig2}{\bf c}, Movie S4,S5). The mean speed, $\overline{u}$, increased by an order of magnitude as the ATP concentration is increased by two orders of magnitude (Fig.~\ref{fig2}{\bf c}). Dividing the average speed of the unsheared network with the characteristic shear rate results in an ATP-independent length scale $\overline{u}/\dot\gamma_c\sim 12$~\si{\micro}m, providing a connection between internal speeds and external shear rates (Fig. \ref{fig2}{\bf c:}Inset). We interpret this length scale as the average distance that microtubule filaments of opposite polarity travel prior to losing mechanical contact. Using this length scale we map internal speeds onto effective active rates. Assuming that the distribution of MT speeds is proportional to the distribution of extensile sliding rates, $P(\dot\epsilon) \propto P(u)$, we define $\dot\epsilon = u /(\overline{u} / \dot{\gamma}_c)$ using the ATP-independent proportionality constant. The fraction of slowly sliding elements in the network is estimated from the cumulative distribution function of the extensile sliding rates $C(\dot{\epsilon}) = \int_{0}^{\dot\epsilon} P(\tilde{\dot\epsilon}) \, d\tilde{\dot\epsilon}$ (Fig.~\ref{fig2}{\bf d}). The numerical value of $C(\dot{\epsilon})$ provides the fraction of solid-like elements when evaluated at $\dot\epsilon = \dot\gamma$. For a given ATP concentration, we denote this fraction at a particular shear rate as $C(\dot\gamma)$  (Fig.~\ref{fig2}{\bf d}). With increasing ATP concentration, the cumulative distribution shift to higher strain rates (Fig \ref{fig2}{\bf e}).

\begin{figure}[ht]
\centerline{\includegraphics[width=.5\textwidth]{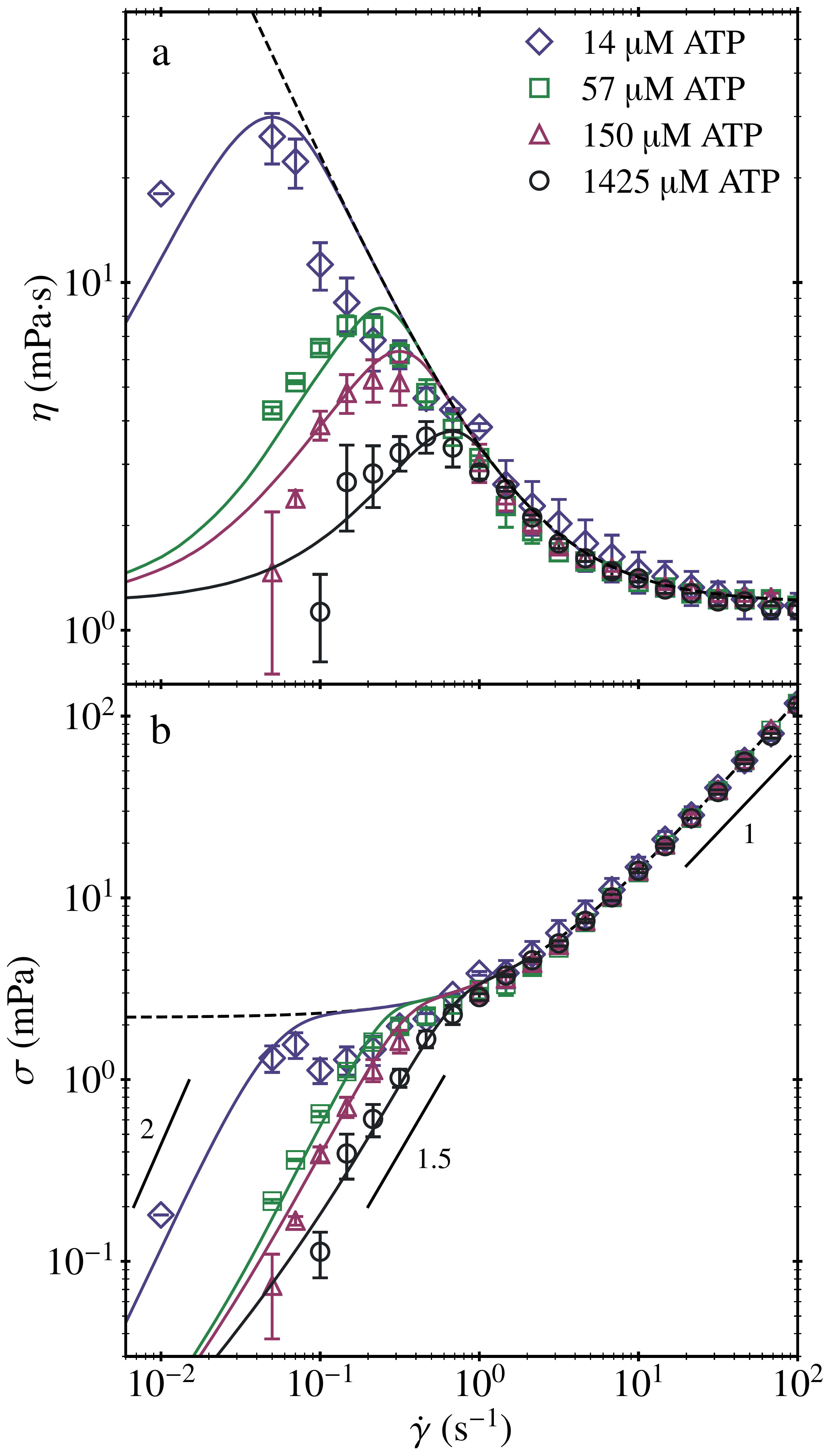}}
\caption{{\bf Modified Bingham plastic model captures rheology of active gels:} Rheological measurements (open symbols) for {\bf a} viscosity and {\bf b} stress. Dashed lines represent the Bingham plastic yield-stress model. The solid lines though the data points are from the modified Bingham model given in Eqs.~\ref{eq:stressModel},\ref{eq:viscModel}, while the labeled lines in {\bf b} provide the power law slopes to guide the eye. At the highest shear rates, the Newtonian plateu must have a $\sigma \sim \dot\gamma $, while at low shear rates, $\sigma\sim\dot\gamma^{1.5-2.0}$.  \label{fig3}}
\end{figure}

The above measurements can explain rheology of active gels can be quantified by extending the Bingham plastic, a model that describes the total stress of shear-yielding materials by a sum of the viscous stress and the yield stress: $\sigma = \eta_s \dot{\gamma} + \sigma_y$\cite{RN1097}. Below $\dot\gamma_c$ for all ATP concentrations, the fraction of solid-like elements determines the rheological response by elastically resisting the shear. Unlike conventional {\em in vitro} biopolymer networks, where the network architecture can be permanently altered when deformed above the yield strain, motor activity provides a mechanism for the continuous reformation of yielded solid-like elements. These solid-like elements contribute to the measured stress at the yield stress of the static gel, $\sigma_y$. Therefore, we modify the Bingham model to incorporate the fraction of slow, solid-like elements within the sheared network by using $C(\dot{\gamma})$ as a coefficient for $\sigma_y$. This preserves the form of the rheological framework while directly accounting for the ATP-dependent fraction of solid-like elements we write:
\begin{equation}
\label{eq:stressModel}
\sigma = \eta_s \dot{\gamma} + C(\dot{\gamma}) \, \sigma_y.
\end{equation}
\begin{equation}
\label{eq:viscModel}
\eta = \eta_s + C(\dot{\gamma})\,\sigma_y/\dot{\gamma}.
\end{equation}
The proposed model has two parameters, the high-shear viscosity, $\eta_s = 1.2$ mPa$\cdot$s, and the yield-stress, $\sigma_y = 2.2$ mPa, both of which were independently measured as described previously (Figs.\ref{fig1}{\bf d,e}). Using these values the model quantitatively describes the rheology of active MT networks for all applied shear rates and ATP concentrations (Fig.~\ref{fig3}{\bf a,b}). Specifically, we observe two distinct stress regimes with an ATP-independent scaling, $\sigma \sim \dot\gamma^{1.5-2.0}$ and $\sigma \sim \dot\gamma$; where the transition maps out a set of points that lie along a plateau determined by the static gel yield-stress. Solid-like elements, therefore, behave like a static gel, and above $\dot\gamma_c$, the growth of the shear stress is dominated by the viscous contribution.

To summarize, we have shown that with decreasing ATP concentration active microtubule networks undergo a transition from a fluid to a solid-like state. The location of the transition is dependent on both the intrinsic ATP-dependent dynamics and the externally applied shear rates. Such fluid-to-solid transition is not predicted by existing theoretical models which assume that active stresses monotonically change with ATP concentration. It follows that the extensile microtubule networks represent a distinct class of active matter systems, whose rheological properties are fundamentally different from suspensions of swimmer based micro-organisms that are described by an established theoretical framework\cite{Saintillan2010, Gachelin2013, Lopez2015,PhysRevE.83.041910, PhysRevLett.92.118101,Gachelin2013, RevModPhys.85.1143,Saintillan2018}. In contrast to hydrodynamically interacting swimmers, system spanning active MT-gels have elastic percolating structures. Theory one must account for both the solid state found at low ATP concentration and the active stresses that effectively melt this solid network. A potential way forward is our finding that self-yielding active gels are quantitatively described by a simple phenomenological model comprised of solid- and liquid-like elements, and whose contribution are determined by the relative distributions of internally generated and externally imposed rates. These findings demonstrate the need to extend microscopic models of passively cross-linked networks to include dynamical events that produce structural rearrangements and bond breakage driven by local active stresses\cite{RevModPhys.86.995}.

\begin{addendum}
 \item We thank Jeffery S. Urbach, Peter D. Olmsted, Emanuela Del Gado, Bulbul Chakraborty, Sriram Ramaswamy, Cristina Marchetti, and Ivan Krocker for discussions. D.A.G., C.D. and D.L.B. thank the Templeton Foundation Grant 57392 and Georgetown University for support. Z.D. and J.B. were supported by the U.S. Department of Energy, Office of Basic Energy Sciences, through award DE-SC0010432TDD. We also acknowledge use of the Brandeis biosynthesis facilities supported by NSF-MRSEC-1420382. 
 \item[Competing Interests] The authors declare they have no
competing financial interests.
 \item[Correspondence] Correspondence and requests for materials
should be addressed to D.L.B.\\
(daniel.blair@georgetown.edu).
\end{addendum}

\newpage

\newpage

\newpage

\end{document}


\maketitle

\begin{affiliations}
 \item Department of Physics, Georgetown University, 3700 O Street NW, Washington DC 20057, USA
 \item Institute for Soft Matter Synthesis \& Metrology, Georgetown University, 3700 O Street NW, Washington, DC 20057, USA
 \item Department of Physics, Brandeis University, Waltham, MA 02453, USA 
 \item Department of Physics, University of California Santa Barbara, Santa Barbara, CA 93106, USA
 $^*$These authors contributed equally to this work
\end{affiliations}
\clearpage
\section*{Rheological procedures and calibration}

Rheological measurements were perfomed using an Anton-Paar MCR 702 TwinDrive system in strain-controlled separate-motor-transducer (SMT) mode to decouple the applied rotation from the torque measurement. The motors were connected to a thermal bath to ensure a constant air-bearing temperature of $23^{\circ}$C that minimizes and maintains a constant internal friction. Experiments were performd at room temperature ($T \approx 21^{\circ}$C).

\begin{figure}[ht]
\centerline{\includegraphics[width=\textwidth]{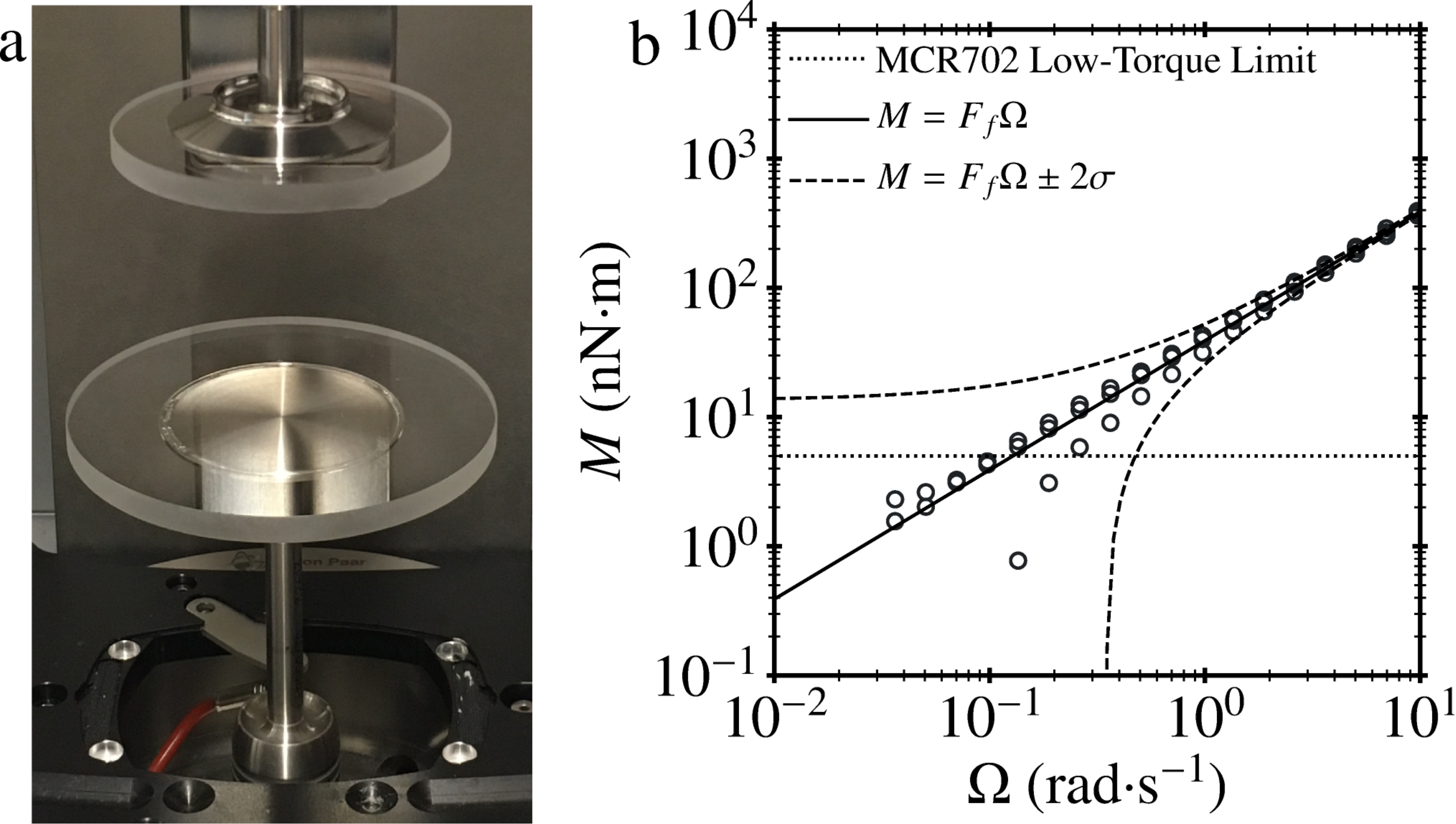}}
\caption{Custom geometry for low viscosity measurements. {\bf a} Custom 75 mm parallel plate geometry made from borosilicate optical windows for Anton Paar MCR702 Twin Drive. {\bf b} Residual torque measurements to characterize the low-torque limit of our custom geometry.  \label{geometry}
}
\end{figure}

We used a custom-made glass parallel-plate measuring system, which was fully interfaced with the rheometer's control software ($D_\mathrm{Top} = 75$~mm and $D_\mathrm{Bottom} = 100$~mm, Fig.~\ref{geometry}{\bf a}). The glass disks have a flatness variation of $\pm2.54$ \si{\micro}m (Edmund Optics). We measured minimal eccentricity ($e = 0.016$), which corresponds to a variation of $\pm5.08$~\si{\micro}m in diameter. Before each use, the glass surface of the measuring system was sonicated with soap and warm water, rinsed with ethanol, and then sonicated in a 0.1 M KOH water solution. Sample evaporation and microtubule migration to the air-liquid interface were mitigated by sealing the sample with solution of light mineral oil (LMO) containing 2\% wt of the silicone emulsifier Abil~EM~90 (Evonik Nutrition \& Care GmbH). We removed the contribution of the sealing oil from our rheological measurements by quantifying the viscosity of water sealed with the LMO solution at various shear rates ($\eta_{H_2O+LMO} \approx 1.7$~mPa$\cdot$s. The measurement gap was kept constant at 280 \si{\micro}m, with a maximum measured variation of 3.81 \si{\micro}m (1.36\%).

The instrument has a manufacturer-stated low-torque limit of 5~nNm in steady shear and 1~nNm in oscillation. Before using the custom measuring system, we ensured inertia and torque compatibility with the rheometer control software. This allows for the use of built-in rotational mapping and motor adjustment tools, which precisely measures the contribution to the torque from internal friction as a function of angular position. We performed residual torque measurements without sample to estimate the internal friction using our custom geometry (Fig.~\ref{geometry}{\bf b}). These calibrations were completed at a gap of 45 mm and an angular speed between $3.64 < \Omega < 9.73 \times$ 10$^{-2}$ rad/s. Torque values were averaged over an integer number of full rotations with a minimum of one rotation at each angular speed.

The velocity-dependent residual torque $M_f = F_f \Omega$, where $F_f$ is the friction factor, was used to correct the measured torque $M_{\mathrm{meas}}$ and obtain actual torque $M = M_{\mathrm{meas}} - F_f$. The resulting linear viscous scaling and the standard deviation are $F_f = \pm 40.1$ nNm/(rad/s) and $2\sigma= \pm 7.8$ nNm, respectively. The uncertainty is in reasonable agreement with the manufacturer specified low-torque limit of 5 nNm in steady shear. This experimental uncertainty corresponds to a viscosity of approximately 0.6 mPa$\cdot$s at a shear rate of 0.01~s$^{-1}$ and 0.06 mPa$\cdot$s at a shear rate of 0.1~s$^{-1}$ for our custom parallel plate geometry.

\begin{figure}[ht]
\centerline{\includegraphics[width=\textwidth]{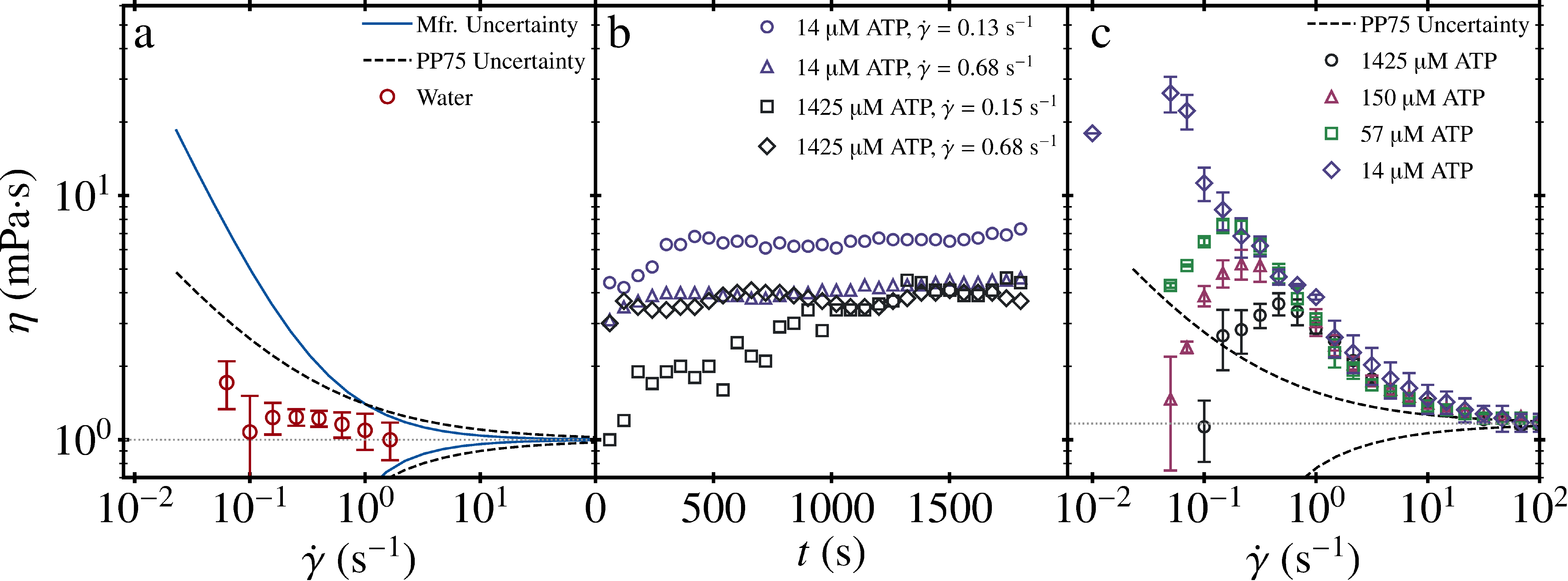}}
\caption{Rheological measurement uncertainty. {\bf a} Measurement uncertainty envelope constructed from viscosity measurements of water. {\bf b} Raw data from rheological measurements of active gels. Data shown in main text is composed of the viscosity data taken from the plateaus of each measurement for a given ATP concentration and applied shear rate. {\bf c} Comparison of uncertainty with our steady-shear rheology for active gels, with most measurements falling well outside the envelope. \label{uncertainty}
}
\end{figure}

We also assessed the torque sensitivity of our custom geometry from the flow curves obtained from a single steady shear rate measurement performed on water (Fig.~\ref{uncertainty}{\bf a}). For each measurement fresh sample was loaded and sealed with light mineral oil containing 2\%wt Abil~EM~90 surfactant (viscosity $\eta_{21^{\circ}C} ~ 40$ mPa$\cdot$s) in order to avoid sample evaporation and minimize the contribution of surface tension to the torque signal. The apparent viscous contribution of the sealing oil, 0.8 mPa$\cdot$s, has been subtracted from all measured viscosity values. With this information we construct an uncertainty envelope for the instrument with our custom geometry:
\begin{equation}
\eta^* = \frac{C_{SS} \left(M_\mathrm{exp} + \Delta M \right)}{\dot{\gamma}} \pm \frac{\Delta \eta}{\eta_N} \eta ,
\end{equation}
where $C_{ss}$ is the measuring-system constant used as a conversion factor between torque $M$ and shear stress, $M_{\mathrm{exp}}$ is the expected torque necessary to shear the sample at the set shear rate $\dot{\gamma}$, $\Delta \eta/\eta_N$ is the uncertainty of the viscosity $\eta$ due to sample inhomogeneity, temperature fluctuations, or small discrepancies in sample filling. $\Delta M$ is the uncertainty of the torque due to friction of the drive unit and bearings. While the term is dominant at high shear rates, the impact of $\Delta M$ increases at low shear rates because  $M_\mathrm{exp}$ decreases. Hence, $\Delta M$ is a key parameter for the sensitivity of the measuring device, especially at low shear rates, where the resulting torque is quite low. Our steady shear rate experiments with water demonstrate a significantly lower instrument uncertainty envelope with $\Delta M = 7$ nN$\cdot$m, compared to the manufacturer's conservative estimate of ($\Delta M = 33$ nN$\cdot$m), due to our custom glass PP75 measuring system's increased surface area. 

We allow steady-shear viscosity measurements to stabilize for a minimum 1000 s (Fig.~\ref{uncertainty}{\bf b}). The resulting shear viscosities of our active microtubule suspensions for most cases fall outside the uncertainty envelope of the device over a range of accessible applied shear rates (Fig.~\ref{uncertainty}{\bf c}). This confirms we have a reasonable signal to noise ratio, even at very low shear rates.

\begin{figure}[ht]
\centerline{\includegraphics[width=\textwidth]{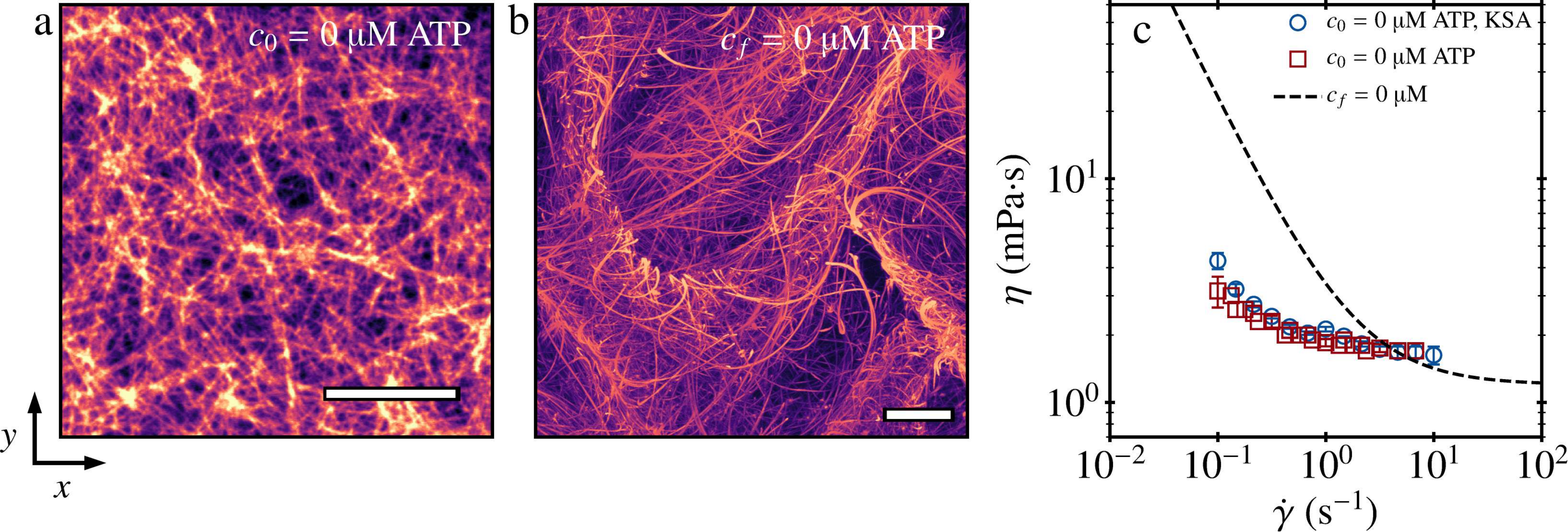}} 
\caption{Rheology of two zero-ATP states: {\bf a-b} Confocal fluorescence images of active network in the ({\em x,y})-plane  $c_0 = 0$ and {\bf b} as $c_{\rm{ATP}} \to 0$, representing $100$~\si{\micro}m  maximum projections. {\bf c} Continuous shear rheology for MT-gels with zero ATP. System with zero initial ATP (blue circles), and the system with zero initial ATP and KSA (orange squares) have identical flow curves and have the morphology shown in {\bf a}. These curves show that with no motor activity, the system is slightly shear thinning, but rapidly approaches the viscosity of the solvent ($\eta\sim 10^{0}$ mPa$\cdot$s). The dashed line is the flow curve for a system that reached a final ATP=0 state through consumption of ATP and has the morphology of the network shown in {\bf b}. The functional form of this curve is the same as the Bingham plastic model used to describe our material with $C=1$.  Scale bars are 50~\si{\micro}m. \label{noATP}
}
\end{figure}

Finally, we examined the case in which no ATP was added to the MT suspension (Fig.~\ref{noATP}a). In this case, the MTs form a network with a different structure. The kinesin motor clusters are unable to traverse along MT-bundles in the absence of ATP. The morphology of this network is fundamentally different from the percolated polarity-sorted structure formed by initiating the binding of motors to MTs with ATP (Fig.~\ref{noATP}{\bf b}). This produces a low-viscosity shear-thinning suspension (Fig.~\ref{noATP}{\bf c}) with a viscosity that is lower than the active gels containing ATP.

\section*{MT dynamics under shear}
\begin{figure}[ht]
\centerline{\includegraphics[width=0.8\textwidth]{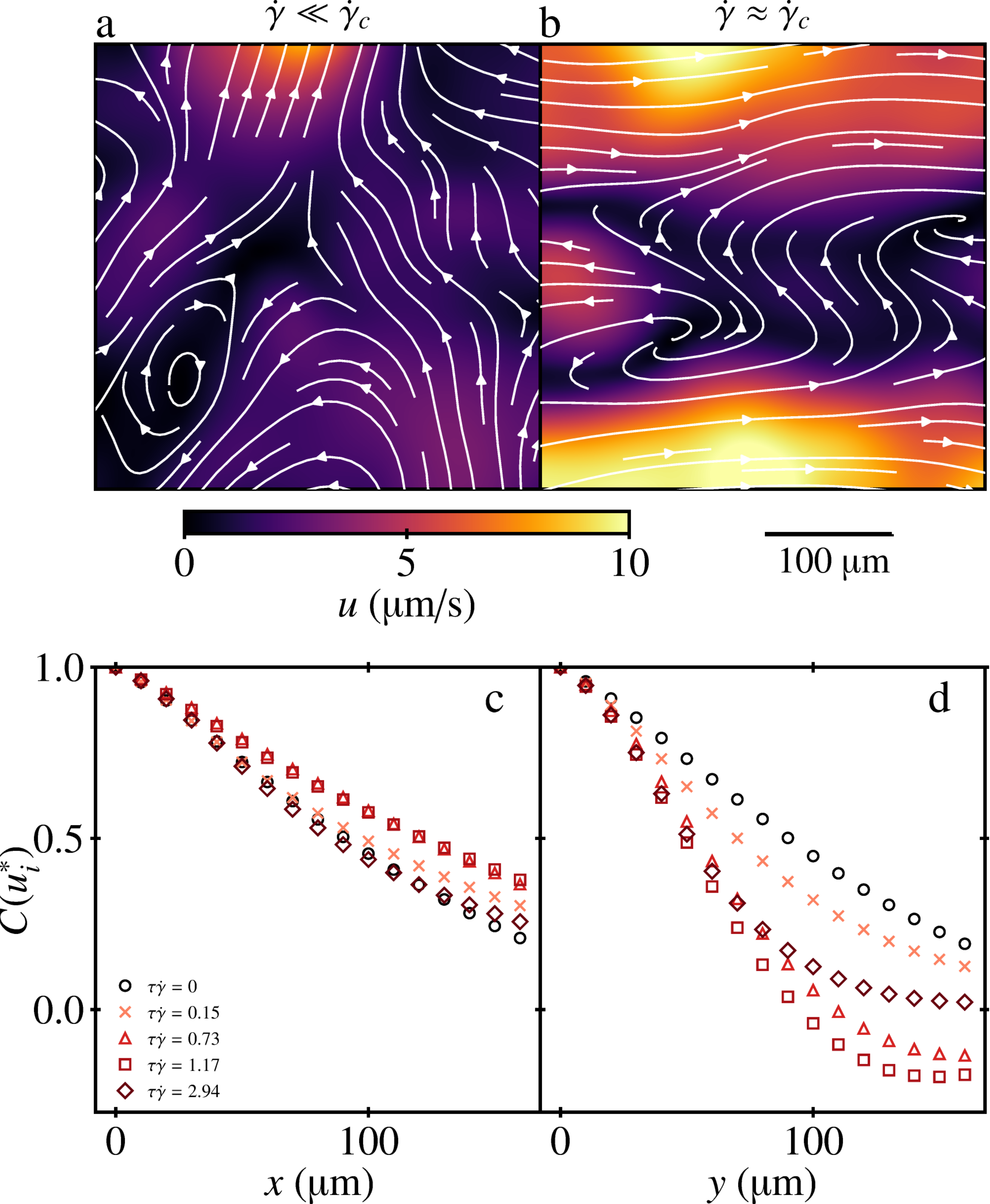}}
\caption{Microtubule dynamics under shear. {\bf a-b} Shear rates smaller and comparable to the internal strain rates produce significantly different non-affine flow fields. Shear rates close to the critical shear rate producing maximal alignment of the non-affine velocity. {\bf c-d}Velocity autocorrelation functions in the shear and vorticity directions show non-monotonic increase and decrease in characteristic length scales, respectively. \label{sheared}
}
\end{figure}

Using particle imaging velocimetry, we measured the flow fields of active gels, which are being sheared in a confocal rheometer under the same experimental condition as our rheological measurements. We then subtract the affine flow fields to remove the sheared deformation from the velocity measurements. This reveals a flow field that reflects the combination of MT activity, buckling, and elastic instabilities. Comparing two flow fields acquired under minimal and strong external shear, reveals that the latter significantly aligns the non-affine flow components in the shear direction, largely eliminating isotropy observed in the limit of no shear (Fig.~\ref{sheared}{\bf{a,b}}). To quantify the alignment of the microtubule activity, we measure the velocity autocorrelation function in the shear ($x$) and vorticity ($y$) directions (Fig.~\ref{sheared}{\bf{c,d}}). The correlation length is non-monotonic in both directions, with shear first enhancing the correlation in the $x-$direction and diminishing the correlation in the $y-$direction. However, once enough external shear is applied and the MT network begins to break, these modifications in the velocity correlations revert towards the isotropic case in the absence of shear due to out-of-plane buckling. We define an approximate correlation length scale as the point at which the correlation has decayed by $1/2$, and the anisotropy as the ratio of those lengthscales $\Lambda = \lambda_x/\lambda_y$.

\section*{Oscillatory rheology}
In addition to performing oscillatory measurements to determine the behavior of the network moduli and yield stress as $c_{ATP}$ was consumed, we also performed a frequency sweep at 1\% strain in an effort to further characterize the relative viscous and elastic contributions of a non-Newtonian material response over a range of timescales. We chose to perform our oscillatory measurements with the lowest ATP concentration (14~\si{\micro}M) because its resistance to flow and therefore torque signal in the rheometer would be largest at low frequency. 

The rheometer geometry (Fig.~\ref{geometry}{\bf a}) and sensitivity sets limits on the ranges for a frequency sweep. The large tool radius enhances our sensitivity and allows for proper surface preparation. However, the inertia of our 75~mm parallel plate geometry is very large. The magnitude of the potential inertial signal for our rheometer geometry in terms of the material moduli is given by:
\begin{equation}
G = \frac{I F_{\tau}}{F_\gamma},
\end{equation}
where $I$ is the moment of inertia, $F_{\tau}$ is the geometric conversion from torque to stress, and $F_{\gamma}$ is the geometric conversion from rotation rate to shear rate\cite{Ewoldt2015}. For confident results, the material torque must be larger than the inertial torque, which for our geometry and sample, only occurs below $\omega \approx 2$. Furthermore, we can estimate the minimum moduli for which we can make a reliable measurement:
\begin{equation}
G = \frac{F_{\gamma}M_{\mathrm{min}}}{\gamma_0},
\end{equation}
where $M_{\mathrm{min}}$ is the minimum torque for the rheometer's mode of operation and $\gamma_0$ is the strain amplitude\cite{Ewoldt2015}. The rheometer manufacturer estimates that the rheometer is approximately five times more sensitive in oscillatory more than steady shear mode. Given the uncertainty we measured in Fig.~\ref{geometry}{\bf b} of $\Delta M =7.8$ nNm, we approximate our minimum torque at $M_{\mathrm{min}} \approx 1.6$ nNm, close to the manufacturer specified minimum torque of 1 nNm in oscillatory mode. Unfortunately, after adding this restriction to our frequency sweep, only a narrow band of frequencies ($0.5<\omega<2$ rad/s) are accessible for our geometry and material.

\section*{Active microtubules}
Fluorescently labeled microtubules, stabilized by the non-hydrolysable GTP analogue, GMPCPP, were purified from bovine tubulin as described previously\cite{Sanchez2012, castoldi2003purification}. Truncated kinesin motors with C-terminal biotin labels, containing the first 401 amino acids of the \textit{Drosophila melanogaster} kinesin-1 heavy chain, were expressed in \textit{Escherichia coli} and purified using standard nickel affinity purification\cite{Berliner1995}. To produce motor clusters, these dimeric kinesin motor proteins were bound to streptavidin in a 2:1 molar ratio. Active suspensions were assembled with microtubules, depletant polymer (Pluronic F-127), kinesin motor clusters, an ATP regeneration system (ATP, pyruvate kinase and phosphoenolpyruvate), and anti-photobleaching agents (glucose, glucose oxidase, and catalase). Even in saturating ATP conditions, the kinesin motor clusters hydrolyze ATP more slowly than the ATP regeneration rate. Thus, the ATP concentration is constant until the phosphoenolpyruvate is wholly consumed hours later.

\section*{Particle imaging velocimetry}
Fluorescent microtubule tracking was carried out on an Anton Paar MCR 301 stress-controlled rheometer integrated with a Leica SP5 scanning confocal microscope to approximate the geometry used for rheological measurements\cite{dutta2013development}. Fluorescently-labeled active MT suspensions were loaded into a 25 mm glass parallel plate measuring system with a 200 \si{\micro}m gap. Sample sealing and glass geometry boundaries were handled the same way as the rheology measurements. Fluorescence imaging is performed at a wavelength of 633 nm with a 10x dry objective between 6 and 13.33 frames per second. We used PIV protocols based on the open source OpenPIV project\cite{Taylor2010}. Flow fields were gently smoothed in time and space with a 3D Gaussian kernel to ensure local continuity of the velocity signals during post-processing. Probability and cumulative distributions are binned and normalized functions of the measured speeds and contain at a minimum $3 \times 10^6$ velocity measurements.

\section*{References}

\bibstyle{naturemag}


\newpage
\newpage
\newpage
\newpage
\newpage
\begin{figure}[ht]
\centerline{\includegraphics[width=\textwidth]{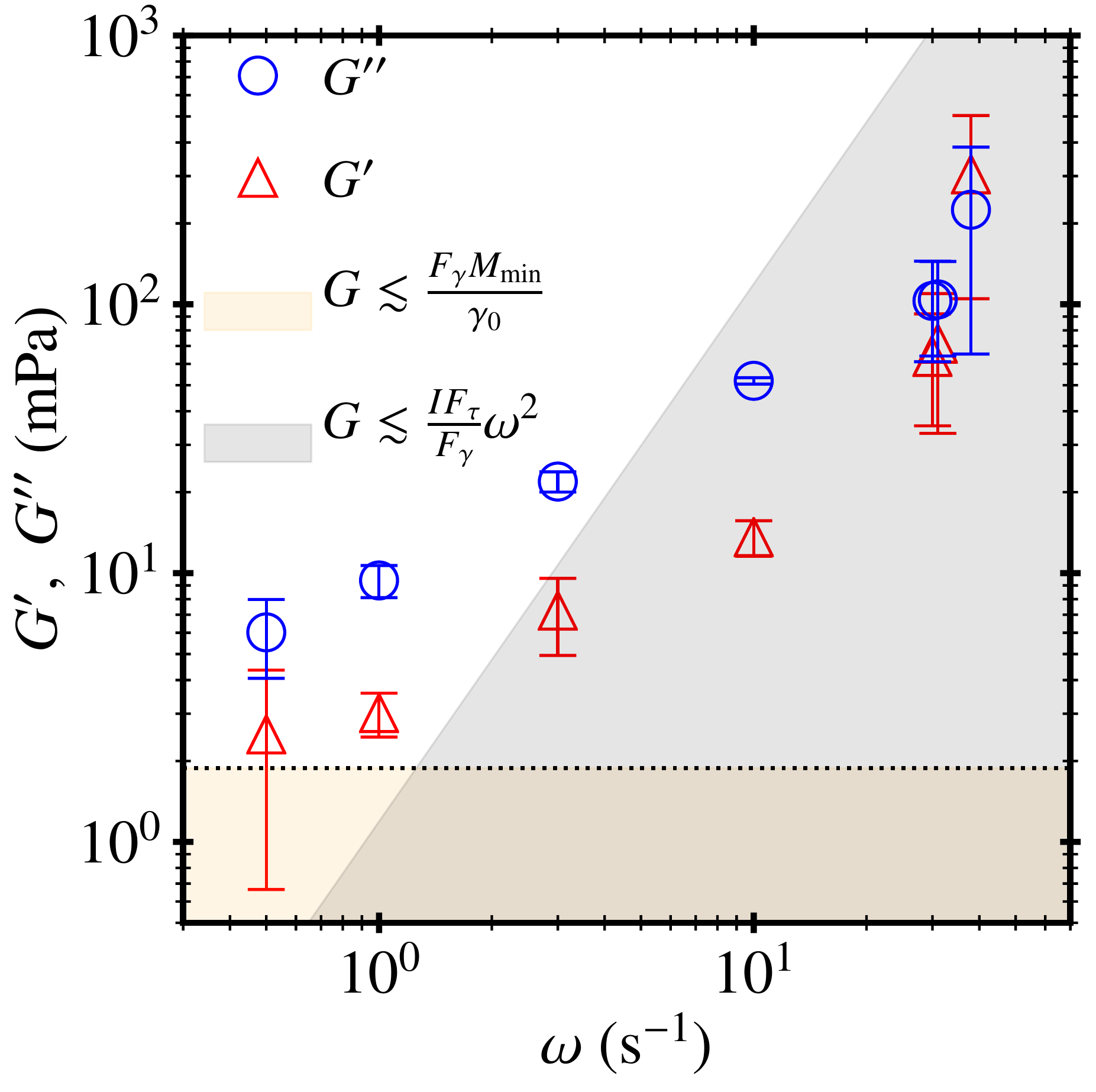}}
\caption{Frequency sweep at 1\% strain. Regions outside of the sensitivity ranges shaded. Grey denotes the range of frequencies and moduli for which the inertial signal is comparable to the material response. The tan region indicates the sensitivity limit for our rheometer derived from our steady shear testing and adapted to oscillatory motion (Fig.~\ref{geometry}{\bf b}) \label{Oscill}
}
\end{figure}
\clearpage

{\hspace{-31pt} Movie 1: Percolated active network: A 3D fluorescent scan of an extensile active gel with saturated $c_{\rm{ATP}}=1425$~\si{\micro}M, demonstrating that the MT network percolates its container while active. Note that this active gel contains a different depletant, polyethylene glycol (PEG), which produces coarser MT bundles more easily visualized in 3D in real time.}

{\hspace{-31pt} Movie 2: Active MTs at lower ATP concentration: Two-dimensional fluorescent confocal images showing the activity of the viscoelastic MT bundles at $c_{\mathrm{ATP}}=57$~\si{\micro}M. The field of view is 300~\si{\micro}m per side.}

{\hspace{-31pt} Movie 3: Active MTs at saturated ATP concentration: Two-dimensional fluorescent confocal images showing the activity of the viscoelastic MT bundles at $c_{\mathrm{ATP}}=1425$~\si{\micro}M. The field of view is 300~\si{\micro}m per side.}

{\hspace{-31pt} Movie 4: MT flow fields at lower ATP concentration: Instantaneous microtubule speeds and streamlines for $c_{\mathrm{ATP}}=57$~\si{\micro}M generated via particle imaging velocimetry on confocal images from Movie 2. Scale bar is 50~\si{\micro}m.}

{\hspace{-31pt} Movie 5: MT flow fields at saturated ATP concentration: Instantaneous microtubule speeds and streamlines for $c_{\mathrm{ATP}}=1425$~\si{\micro}M generated via particle imaging velocimetry on confocal images from Movie 3. Scale bar is 50~\si{\micro}m.}